\documentclass[showpacs,aps,prd,nofootinbib,floatfix,amsmath,amssymb]{revtex4}
\usepackage{graphicx}

\begin{document}


\title{Two-body $Z'$ decays in the minimal $331$ model}
\author{M. A. P\' erez}
\email[E-mail:]{mperez@cinvestav.fis.mx} \affiliation{Departamento
de F\'\i sica, CINVESTAV, Apartado Postal 14-740, 07000, M\'exico,
D. F., M\'exico}
\author{G. Tavares-Velasco}
\email[E-mail:]{gtv@fcfm.buap.mx}
\author{J. J. Toscano}
\email[E-mail:]{jtoscano@fcfm.buap.mx} \affiliation{Facultad de
Ciencias F\'\i sico Matem\'aticas, Benem\'erita Universidad
Aut\'onoma de Puebla, Apartado Postal 1152, Puebla, Pue., M\'
exico}

\date{\today}

\begin{abstract}
The two-body decays of the extra neutral boson $Z_2$ predicted by
the minimal $331$ model are analyzed. At the three-level it can
decay into standard model particles as well as exotic quarks and
the new gauge bosons predicted by the model. The decays into a
lepton pair are strongly suppressed, with $Br(Z_2\to
l^+l^-)\approx 10^{-2}$ and $Br(Z_2\to \bar{\nu}_l \nu)\approx
10^{-3}$. In the bosonic sector, $Z_2$ would decay mainly into a
pair of bilepton gauge bosons, with a branching ratio below the
$0.1$ level. The $Z_2$ boson has thus a leptophobic and
bileptophobic nature and it would decay dominantly into quark
pairs. The anomaly-induced decays $Z_2\to Z_1\gamma$ and $Z_2\to
Z_1Z_1$, which occurs at the one-loop level are studied. It is
found that $Br(Z_2\to Z_1\gamma)\approx 10^{-9}$ and $Br(Z_2\to
Z_1Z_1)\approx 10^{-6}$ at most. As for the $Z_2\to W^+W^-$ and
$Z_2\to Z_1H$ decays, with $H$ a relatively light Higgs boson,
they are induced via $Z'-Z$ mixing. It is obtained that $Br(Z_2\to
W^+W^-)\approx 10^{-2}$ and $Br (Z_2 \to Z_1H)\approx 10^{-5}$. We
also examine the flavor changing neutral current decays $Z_2\to
tc$ and $Z_2\to tu$, which may have branching fractions as large
as $10^{-3}$ and $10^{-5}$, respectively, and thus may be of
phenomenological interest.
\end{abstract}


 \maketitle

\section{Introduction}
\label{int}Extra neutral gauge bosons can arise when the standard
model (SM) group is extended with additional gauge symmetries or
embedded into a larger gauge group. The phenomenology of this
particle has been the subject of significant interest in the
literature \cite{REVIEW}. Current experimental data have been used
to get lower bounds on its mass and the $Z'-Z$  mixing angle
$\theta$. Though it is not possible to obtain model-independent
bounds, current limits from collider \cite{BCOLLIDER} and
precision \cite{PREC} experiments imply that $m_{Z_2}\gtrsim 500$
GeV and $\sin \theta \le 10^{-3}$. However, this situation may be
dramatically different in some models with nonuniversal flavor
gauge interactions, such as topcolor assisted technicolor
\cite{TAT}, noncommuting  extended technicolor \cite{NCET}, or the
ununified standard model \cite{USM}. In this context, it was found
that the extra $Z_2$ boson must be heavier than $1$ TeV
\cite{Chivukula}. Such a gauge boson may be detected at the
Tevatron Run-2. Furthermore, it is expected that the next
generation of colliders will be able to produce a $Z_2$ boson with
a mass up to $5$ TeV \cite{NLC}. It has also been argued
\cite{ARG} that hints of deviations in atomic parity violation
\cite{APV} and the NuTeV experiment \cite{NuE} can be better
explained if the SM is enhanced with an additional $Z_2$ boson.

In this work we are interested in studying some phenomenological
aspects of the extra $Z_2$ boson predicted by the minimal $331$
model \cite{P,Frampton}, which is based on the $SU_C(3)\times
SU_L(3)\times U_X(1)$ gauge group and has some interesting
features \cite{PF1}, such as the possibility of yielding signals
of new physics at the TeV scale. In particular, the effects of
this model on $R_b$ \cite{JAIN} and $t\bar{t}$ production at
hadron colliders \cite{GOMEZ} were analyzed recently. Our main
goal is to investigate the two-body decay channels of the $Z_2$
boson into fermions and gauge bosons. Particular emphasis will be
given to the rare anomaly-induced decay modes $Z_2\to Z_1\gamma$
and $Z_2\to Z_1 Z_2$, where $Z_1$ stands for the SM $Z$ gauge
boson.

In the 331 model the lepton spectrum is the same as in the SM, but
it is arranged in antitriplets of the gauge symmetry $SU_L(3)$.
The quark sector is also arranged in the fundamental
representation of this group, which requires the introduction of
three new quarks. An interesting feature of the model is that
anomalies cancel out when all of the generations are summed over,
which means that the family number must be a multiple of the color
number, which suggest a possible approach to solve the generation
replication problem. In this context, the rare $Z_2\to Z_1\gamma$
and $Z_2 \to Z_1Z_1$ decays are a unique prediction of the model
since they are well defined only if the theory is anomaly free.
Although the decay of a $Z$ boson into two photons or gluons is
the subject of interest in the context of noncommutative quantum
field theories \cite{NONC}, it is worth mentioning that this class
of decays are forbidden in standard field theories by angular
momentum conservation and Bose statistics (Landau-Yang theorem
\cite{YT}).

Along with the $Z_2$ boson, four additional charged gauge bosons
are predicted by the minimal 331 model: two singly charged bosons
$Y^\pm$ and two doubly charged ones $Y^{\pm \pm}$. These gauge
bosons carry two units of lepton number and so have been
classified as bileptons \cite{BILEPTONS}. The new gauge bosons
together with the exotic quarks are endowed with mass at the first
stage of spontaneous symmetry breaking (SSB), when $SU_L(3)\times
U_X(1)$ is broken into $SU_L(2)\times U_Y(1)$ \cite{NG1,VIET}.
Since $SU_L(2)$ is completely embedded into $SU_L(3)$, the
couplings between the SM and the extra gauge bosons are determined
by the coupling constant $g$ associated with the $SU_L(2)$ group
and the weak angle $\theta_W$ \cite{VIET,TT1}. Very interestingly,
the mass of the new gauge bosons are bounded from above due to a
theoretical constraint which yields $\sin\theta^2_W=s^2_W \le 1/4$
\cite{Frampton,NG2}. The fact that the value of $s^2_W$ is very
close to $1/4$ at the $m_{Z_1}$ scale leads to an upper bound on
the scale associated with the first stage of SSB, which translates
directly into a bound on the $Z_2$ mass \cite{Frampton,NG2}. It
was found that the condition $s^2_W(m_{Z_2})\leq 1/4$ can be
translated into the upper bound $m_{Z_2}\leq 3.1$ TeV \cite{NG2}.
Taking into account this bound and the SSB hierarchy determined by
the minimal Higgs sector of the model, it was found that the
bilepton masses cannot be heavier than $m_{Z_2}/2$ \cite{NG2},
thereby allowing the $Z_2\to Y^\pm Y^\mp$, and $Z_2\to Y^{\pm
\pm}Y^{\mp \mp}$ decay modes. The latter is very interesting due
to its distinctive signature $Y^{\pm \pm}\to 2\,l^\pm$.

The upper limit imposed on $s^2_W$ has also some interesting
dynamical consequences. It turns out that the couplings of the
$Z_2$ boson to a lepton pair are given by the $X$ and $T^8$
generators of $U_X(1)$ and $SU_L(3)$, respectively. $X$ is
multiplied by the relatively large factor $1/\sqrt{1-4s^2_W}$ and
$T^8$ by its inverse. The key point is that the coupling of $Z_2$
to a lepton pair is determined exclusively by $T^8$ because the
leptons are in an $X=0$ representation. This is to be contrasted
with the couplings to quark pairs, which are dominated by a term
proportional to $1/(1-4s^2_w)$ as they have a quantum number $X
\neq 0$. Consequently, the extra $Z$ boson of this model is
leptophobic \cite{GOMEZ}. It is important to stress that a similar
situation arises in the case of the couplings to bileptons as they
do not carry quantum number $X$. As will be seen below, these
couplings turn out to be proportional to $\sqrt{1-4s^2_W}$, and
thus the $Z_2$ boson is also bileptophobic. It is clear then that
the leptonic and bileptonic decay modes of the $Z_2$ boson are
expected to be rather suppressed as compared to the quark pair
decays. The fact that the $X$ quantum number of the
third-generation quarks differs from that of the first two
generations leads to flavor changing neutral current (FCNC)
effects mediated by the $Z_2$ boson. Since the $Z_2tq'$ $(q'=u,c)$
couplings are generated at the tree level, it is worth discussing
these rare FCNC transitions of the $Z_2$ boson.

Another decay mode which is of phenomenological interest is
$Z_2\to W^- W^+$. Though the $Z_2$ boson cannot couple directly to
SM bosonic particles, the $Z_2\to W^-W^+$ decay can be induced at
the tree level through the $Z'-Z$ mixing. A similar situation
arises in the case of the $Z_2\to Z_1H$ decay, with $H$ a
relatively light Higgs boson. In this case, a SM-like coupling
$HZ_1Z_1$ is expected and thus the $Z_2\to Z_1H$ mode can be
induced via the $Z'-Z$ mixing. The transition probabilities for
the $Z_2\to W^-W^+$ and $Z_2\to Z_1H$ decays are proportional to
$\sin^2\theta$ and thus one would expect that they are negligibly
small. However, the longitudinal components of the $W$ and $Z_1$
bosons give rise to the factors $(m_{Z_2}/m_W)^4$ and
$(m_{Z_2}/m_{Z_1})^2$, which might compensate the suppression
effect coming from the insertion of the mixing angle.  As  will be
seen below, the $Z_2\to W^+W^-$ decay can be significantly
enhanced if $m_{Z_2}\gg m_{Z_1}$. Though the widths for these
decays do not depend on specific details of the model as they are
determined entirely by the SM couplings and the $\theta$ mixing
angle, their branching ratios are model-dependent indeed.

The rest of the paper is organized as follows. A brief review of
the minimal $331$ model is presented in Sec. \ref{model},
including all the Feynman rules necessary for our calculation.
Sec. \ref{cal} is devoted to the calculation of the two-body
decays $Z_2\to Z_1\gamma$, $Z_1Z_1$, $W^+W^-$, $Y^+Y^-$,
$Y^{++}Y^{--}$, and $\bar{f}f$, included the rare one $Z_2\to
tq'$. Special emphasis is given to the anomaly-induced $Z_2\to
Z_1\gamma$ and $Z_2\to Z_1Z_1$ modes. In Sec. \ref{br} we discuss
the branching fractions associated with each decay mode. Finally,
our conclusions are presented in Sec. \ref{c}.

\section{The minimal 331 model}
\label{model}The $SU_C(3)\times SU_L(3)\times U_X(1)$ model has
been discussed to some extent in the literature
\cite{P,Frampton,PF1,NG2,VIET,TT1}. We will only focus on those
features which are relevant for the present discussion. In this
model the gauge interactions are non flavor-universal since
fermion generations are represented differently  under the
$SU_L(3)$ group. The leptons are accommodated as antitriplets of
$SU_L(3)$:
\begin{equation}
L_{1,2,3}=\left( \begin{array}{ccc}e \\ \nu_e \\e^c
\end{array}\right), \ \
\left( \begin{array}{ccc}\mu \\ \nu_\mu \\ \mu^c
\end{array}\right), \ \
\left( \begin{array}{ccc}\tau \\ \nu_\tau \\ \tau^c
\end{array}\right): \ \ (1,3^*,0).
\end{equation}
Notice that the leptons do not carry $X$ quantum numbers.  In
order to cancel the $SU_L(3)$ anomaly, the same number of fermion
triplets and antitriplets is necessary. This requires to arrange
two quark generations as triplets and the other one as an
antitriplet. It is customary to choose the third generation as the
one transforming as a triplet in order to distinguish the new
dynamic effects in the physics of the quark top from that of the
lighter generations. Accordingly, the three generations are
specified as follows:

\begin{equation}
Q_{1,2}=\left( \begin{array}{ccc}u \\ d \\ D
\end{array}\right), \ \
\left( \begin{array}{ccc}c \\ s \\ S
\end{array}\right): \ \ (3,3,-1/3), \ \ \ Q_3=\left( \begin{array}{ccc}b \\ t \\ T
\end{array}\right): \ \ (3,3^*,2/3),
\end{equation}

\begin{equation}
 d^c, \ \ s^c, \ \ b^c: \ \ (3^*,1,1/3), \ \ \ D^c, \ \ S^c: \ \
(3^*,1,4/3),
\end{equation}

\begin{equation}
u^c, \ \ c^c, \ \ t^c: \ \ (3^*,1,-2/3), \ \ \ T^c: \ \
(3^*,1,-5/3).
\end{equation}
where $D$, $S$, and $T$ are exotic quarks with electric charge
$-4/3$, $-4/3$, and $5/3$, respectively.

The Higgs sector is comprised by three triplets and one sextet of
$SU_L(3)$:
\begin{equation}
\phi_Y=\left( \begin{array}{ccc}\Phi_Y \\ \phi^0
\end{array}\right): \ \ (1,3,1),\ \ \  \phi_1=\left( \begin{array}{ccc}\Phi_1 \\ \delta^-
\end{array}\right): \ \ (1,3,0),\ \ \
\phi_2=\left( \begin{array}{ccc}\widetilde{\Phi}_2 \\ \rho^{--}
\end{array}\right): \ \ (1,3,-1),
\end{equation}

\begin{equation}
H=\left( \begin{array}{ccc}T & \widetilde{\Phi}_3/\sqrt{2} \\
\widetilde{\Phi}^T_3/\sqrt{2} & \eta^{--}
\end{array}\right): \ \ (1,6,0).
\end{equation}
To break $SU_L(3)\times U_X(1)$ into $SU_L(2)\times U_Y(1)$, only
the $\phi_Y$ scalar triplet of $SU_L(3)$ is required. The
hypercharge is identified as a linear combination of the broken
generators $T^8$ and $X$:
$Y=\sqrt{3}(\lambda^8+\sqrt{2}X\lambda^9)$, with $\lambda^8$ a
Gell-Mann matrix and $\lambda^9=\sqrt{2/3}\,{\rm diag}\,(1,1,1)$.
The next stage of SSB occurs at the Fermi scale and is achieved by
the two triplets $\phi_1$ and $\phi_2$. The sextet $H$ is
necessary to provide realistic masses for the leptons \cite{FHPP}.
In these expressions $\Phi_Y$, $\Phi_1$,
$\widetilde{\Phi}_2=i\sigma^2\Phi_2^*$, and $\Phi_3$ are all
doublets of $SU_L(2)$ with hypercharge $3$, $1$, $1$, and $1$,
respectively. On the other hand, $T$ is a $SU_L(2)$ triplet with
$Y=+2$, whereas $\delta^-$, $\rho^{--}$, and $\eta^{--}$ are all
singlets of $SU_L(2)$ with hypercharge $-2,-4$, and $+4$,
respectively \cite{NG1,TT1}. The extra $Z_2$ boson, the bileptons
and the exotic quarks get masses at the first stage of SSB through
the vacuum expectation value $<\phi_Y>_0=(0,0,u/\sqrt{2})$. The
bileptons form an $SU_L(2)$ doublet with hypercharge $+3$. The
spectrum of physical gauge particles is the following. The charged
gauge bosons are given by
\begin{eqnarray}
&&Y^{++}_\mu=\frac{1}{\sqrt{2}}(A^4_\mu-iA^5_\mu),\\
&&Y^+_\mu=\frac{1}{\sqrt{2}}(A^5_\mu-iA^7_\mu), \\
&&W^+_\mu=\frac{1}{\sqrt{2}}(A^1_\mu-iA^2_\mu).
\end{eqnarray}
with $m^2_{Y^{++}}=g^2/4(u^2+v^2_2+3v^2_3)$,
$m^2_{Y^+}=g^2/4(u^2+v^2_1+v^2_3)$, and
$m^2_W=g^2/4(v^2_1+v^2_2+v^2_3)$. The hierarchy of the SSB yields
a splitting between the bilepton masses given by
$|m^2_{Y^+}-m^2_{Y^{++}}|\leq 3m^2_W$.

In the neutral sector, the gauge fields $(A^3,A^8,X)$ define three
mass eigenstates $(A,Z_1,Z_2)$ via the following rotation
\begin{equation}
\left( \begin{array}{ccc} A_\mu \\ Z_\mu \\ Z'_\mu \end{array}
\right)=\left( \begin{array}{ccc} s_W & \sqrt{3}s_W &
\sqrt{1-4s^2_W} \\ c_W & -\sqrt{3}s_Wt_W & -t_W\sqrt{1-4s^2_W}
\\ 0 & -\frac{\sqrt{1-4s^2_W}}{c_W} & \sqrt{3}t_W \end{array} \right) \left(
\begin{array}{ccc} A^3_\mu \\ A^8_\mu \\ X_\mu \end{array}
\right),
\end{equation}
with
\begin{equation}
\left( \begin{array}{ccc} Z_{1\mu} \\ Z_{2\mu}\end{array}
\right)=\left(
\begin{array}{ccc} \cos\theta & -\sin\theta  \\
\sin\theta & \ \ \cos\theta
\\ \end{array} \right) \left(
\begin{array}{ccc} Z_\mu \\ Z'_\mu \end{array}
\right),
\end{equation}
where the mixing angle is
\begin{equation}
\sin^2\theta=\frac{m^2_Z-m^2_{Z_1}}{m^2_{Z_2}-m^2_{Z_1}}.
\end{equation}
with $m^2_Z=m^2_W/c^2_W$ and $Z_1$ standing for the SM $Z$ boson.

On the other hand, by matching the gauge coupling constants at the
first stage of SSB, it is found that
\begin{equation}
\frac{g^2_X}{g^2}=\frac{6s^2_W(m_{Z'})}{1-4s^2_W(m_{Z'})}
\end{equation}
which means that $s^2_W(m_{Z'})$ has to be smaller than $1/4$. It
was found that this condition implies that the extra $Z$ boson
cannot be heavier than $3.1$ TeV \cite{NG2}. From this result and
the symmetry-breaking hierarchy $u \gg v_1,v_2,v_3$, it is
inferred that the bileptons have masses smaller than $m_{Z_2}/2$
\cite{NG2}. In this way, all the new gauge boson masses are
bounded from above. As a consequence, the $Z_2\to
Y^+Y^-,Y^{++}Y^{--}$ decays would always be kinematically allowed.

Some comments concerning the Yang-Mills sector are relevant for
the subsequent discussion. Besides the sector associated with the
electroweak group, the Yang-Mills Lagrangian associated with
$SU_L(3)\times U_X(1)$ gives rise to two new terms, which can be
written in a manifest $SU_L(2)\times U_Y(1)$-invariant form
\cite{TT1}:
\begin{eqnarray}
{\cal L}_{SMNP}&=&-\frac{1}{2}(D_\mu Y_\nu -D_\nu
Y_\mu)^\dag(D^\mu Y^\nu-D^\nu Y^\mu)-Y^{\dag \mu}(ig{\bf W}_{\mu
\nu}+ig'{\bf B}_{\mu \nu})Y^\nu \nonumber \\
&-&\frac{i\sqrt{3}g\sqrt{1-4s^2_W}}{2c_W}Z'_\mu[Y^\dag_\nu (D^\mu
Y^\nu-D^\nu Y^\mu)-(D^\mu Y^\nu-D^\nu Y^\mu)^\dag Y_\nu],
\end{eqnarray}
\begin{eqnarray}
{\cal L}_{NP}&=& -\frac{1}{4}Z'_{\mu \nu}Z'^{\mu
\nu}-\frac{\sqrt{3}g\sqrt{1-4s^2_W}}{2c_W}Z'_{\mu \nu}Y^{\dag
\mu}Y^\nu -\frac{3g^2(1-4s^2_W)}{4c^2_W}Z'_\mu Y^\dag_\nu(Z'^\mu
Y^\nu-Z'^\nu Y^\mu)\nonumber \\
&+&\frac{g^2}{2}(Y^\dag_\mu \frac{\sigma^i}{2}Y_\nu)(Y^{\dag
\mu}\frac{\sigma^i}{2}Y^\nu-Y^{\dag
\nu}\frac{\sigma^i}{2}Y^\mu)+\frac{3g^2}{4}(Y^\dag_\mu
Y_\nu)(Y^{\dag \nu}Y^\nu-Y^{\dag \nu}Y^\mu).
\end{eqnarray}
where ${\bf W}_{\mu \nu}=\sigma^i W^i_{\mu \nu}/2$, ${\bf B}_{\mu
\nu}=YB_{\mu \nu}/2$, and $D_\mu=\partial_\mu -ig{\bf W}_\mu
-ig'{\bf B}_\mu$ is the covariant derivative associated with the
electroweak group, with $Y^\dag_\mu=(Y^{--}_\mu, Y^-_\mu)$ the
$SU_L(2)$ bilepton doublet. The bileptophobic nature of the $Z_2$
boson becomes clear  from these Lagrangians, in which the $Z'YY$
and $Z'Z'YY$ couplings appear multiplied by the factors
$\sqrt{1-4s^2_W}$ and $1-4s^2_W$, respectively. This is a direct
consequence from the fact that the bileptons belong to an $X=0$
representation.

\subsection{Feynman rules}
We will present now the Feynman rules necessary for our
calculation. The current sector is determined by the covariant
derivative associated with $SU_L(3)\times U_X(1)$, which is given
by
\begin{equation}
D_\mu=\partial_\mu-ig\frac{\lambda^a}{2}A^a_\mu-ig_X\frac{\lambda^9}{2}X_\mu,
\end{equation}
in the fundamental representation. $\lambda^a$ ($a=1,...,8$) are
the Gell-Mann matrices and $\lambda^9=\sqrt{2/3}\,{\rm
diag}(1,1,1)$. In terms of physical fields, $D_\mu$ can be written
as
\begin{eqnarray}
D_\mu &=&
\partial_\mu-\frac{ig}{\sqrt{2}}\left(\lambda_{12}W^+_\mu+\lambda_{45}Y^{++}_\mu
+\lambda_{67}Y^+_\mu +{\rm H.c.} \right) -ieQA_\mu \nonumber \\
&&-\frac{ig}{2c_W}\left(c^2_W\lambda^3 -s^2_WY\right)Z_\mu +
\frac{ig}{2c_W}\sqrt{1-4s^2_W}\left(\lambda^8
-\frac{3\sqrt{2}s^2_W}{1-4s^2_W}\lambda^9X\right)Z'_\mu,
\end{eqnarray}
where $\lambda_{ab}=\frac{1}{2}(\lambda^a+i\lambda^b)$. From the
presence of the factor $\sqrt{1-4s^2_W}$ in the term involving the
extra $Z$ boson, the leptophobic nature of $Z'$ becomes clear.

In terms of the $\{A,Z,Z'\}$ fields we can write the neutral
currents as follows:
\begin{equation}
{\cal L}^{NC}=ie\sum_f Q_f(\bar{f}\gamma_\mu
f)A^\mu+\frac{ig}{2c_W}\sum_f\left(\bar{f}\gamma_\mu(g^f_{VZ}-g^f_{AZ}\gamma_5)fZ^\mu
+\bar{f}\gamma_\mu(g^f_{VZ'}-g^f_{AZ'}\gamma_5)fZ'^\mu\right).
\end{equation}
The coefficients $g^f_i$are listed in Table \ref{TABLE}. It turns
out that the couplings of the SM $Z$ boson to exotic quarks are
vector-like. These results can be easily translated into the mass
eigenstates $\{Z_1,Z_2\}$ by means of the relations
\begin{eqnarray}
&&g^f_{VZ_1}=\cos\theta g^f_{VZ}-\sin\theta g^f_{VZ'}, \\
&&g^f_{VZ_2}=\sin\theta g^f_{VZ}+\cos\theta g^f_{VZ'},
\end{eqnarray}

\begin{eqnarray}
&&g^f_{AZ_1}=\cos\theta g^f_{AZ}-\sin\theta g^f_{AZ'}, \\
&&g^f_{AZ_2}=\sin\theta g^f_{AZ}+\cos\theta g^f_{AZ'}.
\end{eqnarray}
However, with the exception of the $Z_2WW$ and $Z_2Z_1H$
couplings, we will ignore $Z'-Z$ mixing effects as they cannot
affect significantly the branching fractions of the main decay
channels.

As far as FCNCs are concerned, they are essentially mediated by
the $Z_2$ boson as the couplings of the SM quarks to $Z_1$ are
proportional to $\sin\theta$ and thus are quite suppressed. These
effects arise solely from the left-handed sector and are a result
from the different $X$ quantum number assignments existing among
the fermion families. We will focus on $Z_2$ transitions involving
the top quark. Thus, the FCNC Lagrangian for the up sector can  be
written as

\begin{equation}
\mathcal{L}_{FCNC}=\frac{g}{2c_W}\left(-\sin\theta
Z^\mu_1+\cos\theta Z^\mu_2\right)\delta_L
V^*_{3a}V_{3b}\overline{U}_a\gamma_\mu P_LU_b,
\end{equation}
where
\begin{equation}
\delta_L=\frac{2}{\sqrt{3}}\,\frac{c^2_W}{\sqrt{1-4s^2_W}},
\end{equation}
$V_{ab}$ is the unitary matrix relating gauge states to mass
eigenstates, and $U_a=u,\,c,\,t$. In the right-handed sector there
is no FCNC as these fermions transform identically.

\begin{table}
\caption{\label{TABLE} Structure of the neutral currents in the
minimal $331$ model. The couplings are given in terms of the
$\{A,Z,Z'\}$ fields.}
\begin{ruledtabular}
\begin{tabular}{llllll}
Fermion & $Q_f$ & $g^f_{VZ}$ & $g^f_{AZ}$ & $g^f_{VZ'}$ &
$g^f_{AZ'}$ \\
\hline $l^-$ & $-1$ & $-\frac{1-4s^2_W}{2}$ & $-\frac{1}{2}$ &
$\frac{\sqrt{3}\sqrt{1-4s^2_W}}{2c^2_W}$ &
$-\frac{\sqrt{1-4s^2_W}}{2\sqrt{3}c^2_W}$ \\
$\nu_l$ & $0$ & $\frac{1}{2}$ & $\frac{1}{2}$ &
$\frac{\sqrt{1-4s^2_W}}{2\sqrt{3}c^2_W}$ &
$\frac{\sqrt{1-4s^2_W}}{2\sqrt{3}c^2_W}$ \\
$u,c$ & $+\frac{2}{3}$ & $\frac{3-8s^2_W}{6}$ & $\frac{1}{2}$ &
$-\frac{1-6s^2_W}{2\sqrt{3}c^2_W\sqrt{1-4s^2_W}}$ &
$-\frac{1+2s^2_W}{2\sqrt{3}c^2_W\sqrt{1-4s^2_W}}$ \\
$d,s$ & $-\frac{1}{3}$ & $ -\frac{3-4s^2_W}{6}$ & $-\frac{1}{2}$
& $ -\frac{1}{2\sqrt{3}c^2_W\sqrt{1-4s^2_W}}$ &
$-\frac{\sqrt{1-4s^2_W}}{2\sqrt{3}c^2_W}$ \\
$D,S$ & $-\frac{4}{3}$ & $\frac{8s^2_W}{3}$ & $0$ &
$\frac{1-9s^2_W}{\sqrt{3}c^2_W\sqrt{1-4s^2_W}}$ &
$\frac{1}{\sqrt{3}\sqrt{1-4s^2_W}}$\\
$b$ & $-\frac{1}{3}$ & $-\frac{3-4s^2_W}{6}$ & $-\frac{1}{2}$ &
$\frac{1-2s^2_W}{2\sqrt{3}c^2_W\sqrt{1-4s^2_W}}$ &
$\frac{1+2s^2_W}{2\sqrt{3}c^2_W\sqrt{1-4s^2_W}}$ \\
$t$ & $+\frac{2}{3}$ & $\frac{3-8s^2_W}{6}$ & $\frac{1}{2}$ &
$\frac{1+4s^2_W}{2\sqrt{3}c^2_W\sqrt{1-4s^2_W}}$ &
$\frac{\sqrt{1-4s^2_W}}{2\sqrt{3}c^2_W}$ \\
$T$ & $+\frac{5}{3}$ & $-\frac{10s^2_W}{3}$ & $0$ &
$-\frac{1-11s^2_W}{\sqrt{3}c^2_W\sqrt{1-4s^2_W}}$ &
$-\frac{1}{\sqrt{3}\sqrt{1-4s^2_W}}$\\
\end{tabular}
\end{ruledtabular}
\end{table}

The $Z_2W^+W^-$, $Z_2Y^+Y^-$, and $Z_2Y^{++}Y^{--}$ couplings
arise from the Lagrangian
\begin{equation}
{\cal
L}_{Z_2VV^\dag}=igc_Wg_{Z_2VV^\dag}\left(Z^\beta_2\left(V^\dag_{\alpha
\beta}V^\alpha-V_{\alpha \beta}V^{\dag \alpha}\right)+Z_{2\alpha
\beta}V^{\dag \alpha}V^\beta\right),
\end{equation}
where $V=W^+,Y^+,Y^{++}$, and

\begin{equation}
g_{Z_2VV^\dag}=\left\{ \begin{array}{ll} -\sin\theta, \ \ \ V=W^+ \\
\frac{\sqrt{3\left(1-4s^2_W\right)}}{2c^2_W}, \ \ \
V=Y^+,Y^{++}\end{array}\right.
\end{equation}
The $Z_{2\lambda}V_\mu(k_1)V^\dag_\nu(k_2)$ vertex can be written
as
\begin{equation}
\label{Gamma_Z2VV} \Gamma^{Z_2VV^\dag}_{\lambda \mu
\nu}(k_1,k_2)=-igc_Wg_{Z_2VV^\dag}\left(2k_{1\nu}g_{\lambda
\mu}-2k_{2\mu}g_{\lambda \nu}+(k_2-k_1)_\lambda g_{\mu
\nu}\right),
\end{equation}
where all the 4-momenta are incoming.

\section{Two-body decays of $Z_2$}
\label{cal} We now turn to the widths for the decay modes $Z_2\to
Z_1\gamma$, $Z_1Z_1$, $YY$, $WW$, $Z_1H$, and $\bar{f}f$. The
$Z_2\to Z_1\gamma,Z_1Z_1$ decays will be discussed more carefully
to clarify some subtleties associated with the triangle anomaly.

\subsection{Anomaly-induced $Z_2\to Z_1\gamma$, $Z_1Z_1$ decays}
As will be seen below,  the amplitudes associated with the $Z_2\to
Z_1\gamma,Z_1Z_1$ decays are model-independent as they are
dictated by gauge invariance and Bose symmetry, respectively,
whereas their magnitude depends on the fermion content of the
theory and the anomaly cancellation mechanism. In the $331$ model,
the couplings of $Z_2$ to fermions are determined by the coupling
constant $g$ and the weak angle only. Also, we recall that the
distinctive feature of the 331 model is that the triangle anomaly
does not cancel out within each generation, but when all the
generations are considered. The $Z_2\to Z_1\gamma$, $Z_1Z_1$
decays have been already studied in the context of a
superstring-inspired $E_6$ model \cite{Dchang}.

\begin{figure}
 \centering
\includegraphics[width=4in]{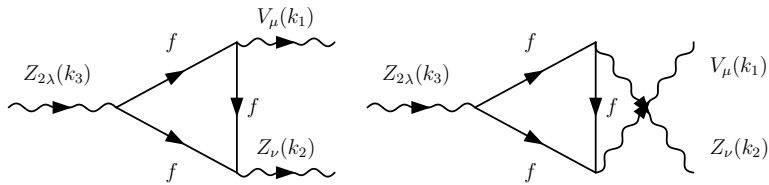}
 \caption{\label{FIG1}Feynman diagrams contributing to the
 anomaly-induced $Z_2\to Z_1V$ decay, with $V=\gamma$ or $Z_1$.}
 \end{figure}

\subsubsection{The decay $Z_2\to Z_1\gamma$}

The $Z_2\to Z_1\gamma$ decay receives contributions from all the
charged fermions via the two triangle diagrams shown in Fig.
\ref{FIG1}, where we also show the notation for the 4-momenta of
the participating particles. The invariant amplitude can be
written as
\begin{equation}
{\cal M}(Z_2\to Z_1\gamma)=\Gamma^{Z_2Z_1\gamma}_{\lambda \mu
\nu}R^{\lambda \mu \nu},
\end{equation}
where $R^{\lambda \mu \nu}$ is a term involving the polarization
vectors associated with the final states:
\begin{equation}
R^{\lambda \mu
\nu}=\epsilon^{\lambda}(k_3,\lambda_3)\epsilon^{*\mu}(k_1,\lambda_1)
\epsilon^{*\nu}(k_2,\lambda_2).
\end{equation}
The $\Gamma^{Z_2Z_1\gamma}_{\lambda \mu \nu}$ tensor is
well-defined only if the theory is anomaly free. In this case, its
form is determined by electromagnetic gauge invariance and must
satisfy the following Ward identity
\begin{equation}
k^\mu_1\Gamma^{Z_2Z_1\gamma}_{\lambda \mu \nu}=0.
\end{equation}
It turns out that the anomaly cancels out after the following
identity is used
\begin{equation}
\sum_f Q_f N^f_C
\left(g^f_{VZ}g^f_{AZ'}+g^f_{VZ'}g^f_{AZ}\right)=0,
\end{equation}
where $Q_f$ is the electric charge in units of the positron charge
and $N^f_C$ is the color index. We stress that the sum runs over
all the charged fermions as the anomaly is not cancelled within
each generation. This identity can be straightforwardly verified
using the values given in Table \ref{TABLE}. Not all the terms
appearing in the calculation are independent. In order to obtain a
gauge invariant result, it is necessary to use Shouten's identity,
which leads to
\begin{equation}
k_{1\lambda}\epsilon_{\mu \nu \alpha
\beta}k^\alpha_1k^\beta_2+k_{1\nu}\epsilon_{\lambda \mu \alpha
\beta}k^\alpha_1k^\beta_2+k_1\cdot k_2\epsilon_{\lambda \mu \nu
\alpha}k^\alpha_1=0,
\end{equation}
We thus obtain
\begin{equation}
\Gamma^{Z_2Z_1\gamma}_{\lambda \mu \nu}=\frac{e\alpha}{4\pi
s^2_{2W}}\frac{m^4_{Z_2}}{(m^2_{Z_2}-m^2_{Z_1})^2}\sum_f
Q_fN^f_C\Big(g^f_{VZ}g^f_{AZ'}+g^f_{VZ'}g^f_{AZ}\Big){\cal A}
P_{\lambda \mu \nu},
\end{equation}
where the loop amplitude ${\cal A}$ is given by
\begin{equation}
{\cal
A}=B_0(1)-B_0(2)-2\left(1-\frac{m^2_{Z_1}}{m^2_{Z_2}}\right)m^2_f
C_0(0,1,2).
\end{equation}
where $B_0(1)=B_0(m^2_{Z_1},m^2_f,m^2_f)$,
$B_0(2)=B_0(m^2_{Z_2},m^2_f,m^2_f)$, and
$C_0(0,1,2)=C_0(0,m^2_{Z_1},m^2_{Z_2},m^2_f,m^2_f,m^2_f)$ are
Passarino-Veltman scalar functions written in the notation of Ref.
\cite{Mertig}. In addition
\begin{equation}
P_{\lambda \mu
\nu}=\frac{1}{m^2_{Z_2}}\left(k_{2\mu}\epsilon_{\lambda \nu \alpha
\beta}k^\alpha_1k^\beta_2-k_{1\nu}\epsilon_{\lambda \mu \alpha
\beta}k^\alpha_1k^\beta_2+k_1\cdot k_2\epsilon_{\lambda \mu \nu
\alpha}k^\alpha_2-\left(k_1\cdot k_2-k^2_2\right)\epsilon_{\lambda
\mu \nu \alpha}k^\alpha_1\right),
\end{equation}
which is manifestly gauge invariant. It is important to notice
that ${\cal A}$ vanishes when ${Z_2}$ and ${Z_1}$ are identical,
which is consistent with the fact that a self-conjugate vector
boson cannot have static electromagnetic properties.

Once the amplitude is squared, the following decay width is
obtained
\begin{equation}
\Gamma(Z_2\to
Z_1\gamma)=\frac{\alpha^3m_{Z_1}}{192\pi^2s^4_{2W}}\left(\frac{m_{Z_1}}{m_{Z_2}}\right)
\left(\frac{m^2_{Z_2}+m^2_{Z_1}}
{m^2_{Z_2}-m^2_{Z_1}}\right)\left|\sum_f
Q_fN^f_C\left(g^f_{VZ}g^f_{AZ'}+g^f_{VZ'}g^f_{AZ}\right){\cal
A}\right|^2.
\end{equation}
We would like to point out that this result is proportional to
$m_{Z_1}$ rather than to $m_{Z_2}$, as one would expect. Notice
also that there is a suppression factor $m_{Z_1}/m_{Z_2}$. These
peculiarities arise from the Lorentz form of the amplitude. In
fact, the squared amplitude is proportional to
\begin{equation}
\left|P_{\lambda \mu \nu}R^{\lambda \mu
\nu}\right|^2=2m^2_{Z_1}\left(1+\frac{m^2_{Z_1}}{m^2_{Z_2}}\right)
\left(1-\frac{m^2_{Z_1}}{m^2_{Z_2}}\right)^2.
\end{equation}
Since $P_{\lambda \mu \nu}$ is determined by electromagnetic gauge
invariance, this result is model-independent and consistent with
Landau-Yang theorem, which requires that $|{\cal M}|^2 =0$ for
$m_{Z_1}= 0$. This means that this decay mode is expected to be
rather suppressed if $Z_2$ is much heavier than $Z_1$.

\subsubsection{The decay $Z_2\to Z_1Z_1$}
Although a vector boson cannot decay into a pair of massless
vector bosons (Landau-Yang theorem), it can decay into a pair of
massive vector bosons. The diagrams which contribute to the
$Z_2\to Z_1Z_1$ decay are shown in Fig. \ref{FIG1}. The
corresponding amplitude can be written as
\begin{equation}
{\cal M}(Z_2\to Z_1Z_1)=\Gamma^{Z_2Z_1Z_1}_{\lambda \mu
\nu}(k_1,k_2)T^{\lambda \mu \nu},
\end{equation}
where
\begin{equation}
T^{\lambda \mu
\nu}=\epsilon^\lambda(k_3,\lambda_3)\epsilon^{*\mu}(k_1,\lambda_1)\epsilon^{*\nu}(k_2,\lambda_2).
\end{equation}
The form of $\Gamma^{Z_2Z_1Z_1}_{\lambda \mu \nu}$ is dictated by
Bose symmetry, which means that it must be symmetric under the
interchange $k_{1\mu}\leftrightarrow k_{2\nu}$,
\begin{equation}
\Gamma^{Z_2Z_1Z_1}_{\lambda \mu
\nu}(k_1,k_2)=+\Gamma^{Z_2Z_1Z_1}_{\lambda \nu \mu}(k_2,k_1).
\end{equation}
In order to cancel the anomaly, it is necessary to use the
identity
\begin{equation}
\sum_f
N^f_C\left\{\left(\left(g^f_{VZ}\right)^2+\left(g^f_{AZ}\right)^2\right)g^f_{AZ'}+2g^f_{VZ}g^f_{VZ'}g^f_{AZ}\right\}=0,
\end{equation}
where the sum runs over all the fermions. The amplitude can be
written as
\begin{equation}
\Gamma^{Z_2Z_1Z_1}_{\lambda \mu \nu}(k_1,k_2)=-\frac{2e\alpha}{\pi
s^3_{sW}}\frac{m^4_{Z_2}}{(m^2_{Z_2}-4m^2_{Z_1})^2}\left(\left({\cal
A}_1+{\cal A}_2+{\cal A}_3\right)P_{1\lambda \mu \nu}+{\cal
A}_1P_{2\lambda \mu \nu}\right),
\end{equation}
where
\begin{eqnarray}
{\cal
A}_1&=&\sum_fN^f_C\left\{\left[\left(g^f_{VZ}\right)^2+\left(g^f_{AZ}\right)^2\right]
g^f_{AZ'}+2g^f_{VZ}g^f_{VZ'}g^f_{AZ}\right
\}\Bigg\{\frac{1}{2}\left(1+\frac{2m^2_{Z_1}}{m^2_{Z_2}}\right)\left(B_0(1)-
B_0(2)\right)\nonumber \\
&&-\left(m^2_{Z_1}\left(1-\frac{m^2_{Z_1}}{m^2_{Z_2}}\right)+
m^2_f\left(1-\frac{4m^2_{Z_1}}{m^2_{Z_2}}\right)\right)
C_0(1,1,2)\Bigg\}, \\
{\cal A}_2&=&4\sum_f
N^f_C\left(g^f_{AZ}\right)^2g^f_{AZ'}\left\{-\frac{2m^2_f}{m^2_{Z_2}}\left(B_0(1)-
B_0(2)\right)+\left(1-\frac{2m^2_{Z_1}}{m^2_{Z_2}}\right)m^2_f
C_0(1,1,2)\right \},\\
{\cal
A}_3&=&4\sum_fN^f_Cg^f_{VZ}g^f_{VZ'}g^f_{AZ}\left(1-\frac{4m^2_{Z_1}}{m^2_{Z_2}}\right)m^2_f
C_0(1,1,2),
\end{eqnarray}
with
$C_0(1,1,2)=C_0(m^2_{Z_1},m^2_{Z_1},m^2_{Z_2},m^2_f,m^2_f,m^2_f)$.
The Lorentz tensors $P_{1\lambda \mu \nu}$ and $P_{2\lambda \mu
\nu}$ read
\begin{eqnarray}
P_{1\lambda \mu
\nu}&=&\frac{1}{2}\left(1-\frac{4m^2_{Z_1}}{m^2_{Z_2}}\right)\epsilon_{\lambda
\mu \nu \alpha}(k_1-k_2)^\alpha, \\
P_{2\lambda \mu
\nu}&=&\frac{1}{m^2_{Z_2}}\Big(k_{1\nu}\epsilon_{\lambda \mu
\alpha \beta}k^\alpha_1k^\beta_2-k_{2\mu}\epsilon_{\lambda \nu
\alpha \beta}k^\alpha_1k^\beta_2\Big),
\end{eqnarray}
which are clearly symmetric under the interchange
$k_{1\mu}\leftrightarrow k_{2\nu}$. When the amplitude is squared,
the contraction of $T^{\lambda \mu \nu}$ with $P_{i\lambda \mu
\nu}$ yields a term proportional to
$m^2_{Z_2}(m^2_{Z_2}/m^2_{Z_1})$, which means that this decay is
essentially determined by the longitudinal component of the $Z_1$
boson, which is to be contrasted with the case of the $Z_2 \to
Z_1\gamma$ mode, which yields a squared amplitude proportional to
$m^2_{Z_1}$. Since $P_{1\lambda \mu \nu}$ and $P_{2\lambda \mu
\nu}$  are determined only by Bose symmetry, this behavior is
model-independent indeed.

The decay width can be written as
\begin{equation}
\Gamma(Z_2\to
Z_1Z_1)=\frac{\alpha^3m_{Z_2}}{24\pi^2s^6_{2W}}\Big(\frac{m_{Z_2}}{m_{Z_1}}\Big)^2
\sqrt{1-\frac{4m^2_{Z_1}}{m^2_{Z_2}}}\left|{\cal A}_1+{\cal A}_2-
\frac{4m^2_{Z_1}}{m^2_{Z_2}-4m^2_{Z_1}}{\cal A}_1\right|^2,
\end{equation}
where we have introduced a factor of $1/2$ to account for two
identical particles in the final state. From this expression it
can be seen that the decay width vanishes when $m_{Z_1}=0$ and
$g^f_{AZ}=0$, which is consistent with Landau-Yang theorem.

\subsection{The decay $Z_2\to V^\dag V$}
We turn now to the analysis of the $Z_2\to V^\dag V$ decays, with
$V=W^-$, $Y^-$, or $Y^{--}$. The respective amplitude can be
written as
\begin{equation}
{\cal M}(Z_2\to VV^\dag)=\Gamma^{Z_2VV^\dag}_{\lambda \mu
\nu}(k_1,k_2)\epsilon^\lambda(k_3,\lambda_3)\epsilon^{*\mu}(k_1,\lambda_1)\epsilon^{*\nu}(k_2,\lambda_2),
\end{equation}
where the $\Gamma^{Z_2VV^\dag}_{\lambda \mu \nu}(k_1,k_2)$ is
given in Eq. (\ref{Gamma_Z2VV}). The longitudinal component of the
$V$ boson gives rise to a term proportional to
$m^4_{Z_2}(m_{Z_2}/m_{V})^4$, which plays a very important role in
the case of the $Z_2\to W^+W^-$ channel. The decay widths for the
$Z_2 \to VV^\dag$ modes read
\begin{equation}
\Gamma(Z_2\to
W^+W^-)=m_{Z_2}\left(\frac{\alpha}{3}\right)\left(\frac{c_W}{s_W}
\right)^2 \sin^2\theta\, f(x_W),
\end{equation}

\begin{equation}
\Gamma(Z_2\to YY^\dag)=m_{Z_2}\frac{\alpha
(1-4s^2_W)}{s^2_{2W}}f(x_Y),
\end{equation}
where $x_V=4m^2_V/m^2_{Z_2}$, and
\begin{equation}
f(x)=\frac{\sqrt{1-x}}{x^2}\left(1+4x-\frac{17}{4}x^2-\frac{3}{4}x^3\right).
\end{equation}
\indent The $Z_2W^+W^-$ vertex has been studied in several
contexts. The production of the $Z_2$ boson followed by  the
$Z_2\to W^+W^-$ decay has been studied within theories in which
the extra $Z$ boson does not couple to the fermions \cite{NANDI}
and also in the context of the $E_8\times E^\prime_8$ superstring
theory \cite{NAJIMA}. The implications of a relatively heavy extra
$Z$ boson on this decay mode within the context of string $E_6$
inspired theories were investigated in Ref. \cite{DESHPANDE}. The
magnitude of the $Z_2\to W^+W^-$ decay width depends crucially on
the size of $\theta$, which is expected to be very small. From
general grounds, one can assume that $\theta \approx \delta
m^2/m^2_{Z_2}\ll 1$, with $\delta m^2$ a model-dependent quantity.
This means that $\theta$ may be small because either $m_{Z_2}\gg
m_{Z_1}$ or $\delta m^2$ is very small. Thus, a relatively light
$Z_2$ boson is still compatible with a very small mixing angle
provided that $\delta m^2$ is small. Most analysis from precision
experiments support this scenario. In fact it has been found that
$\sin \theta \le 10^{-3}$ for $m_{Z_2}\gtrsim 500$ GeV (for an
exception to this behavior see Ref. \cite{Chivukula}). Another
interesting possibility arises when $Z_2$ is very heavy, which
yields a very small mixing angle even if $\delta m^2$ is of the
order of $m^2_{Z_1}$. In this case, $\theta \approx
m^2_{Z_1}/m^2_{Z_2}$, which is very small because $m_{Z_2}\gg
m_{Z_1}$. In this scenario, the suppression effect coming from
$\sin \theta$ would be compensated by the factor $1/x^2_W$, and
thus the $Z_2 \to W^-W^+$ mode might be as important as the
fermionic channels \cite{DESHPANDE}. In general, if $\delta m^2$
is either very small or of order $O(m^2_{Z_1})$, the $Z_2\to
W^+W^+$ mode is favored for $m_{Z_2}\gg m_{Z_1}$. As for the decay
$Z_2\to YY$, it is expected to have a relatively small branching
ratio, mainly due to the bileptophobic character of $Z_2$ and also
to phase space.

\subsection{The decay $Z_2\to Z_1H$}
The minimal Higgs sector of the 331 model has been already studied
in the literature \cite{HSector}. The three triplets and the
sextet lead to 22 physical states: 5 neutral CP-even, 3 neutral
CP-odd, 4 singly charged, and 3 doubly charged. It is expected
that at least one of the neutral CP-even states resembles the
behavior of the SM Higgs boson. In principle, the $Z_2$ boson may
decay into a pair of singly or doubly charged Higgs bosons, but in
order to simplify our analysis we will assume that these channels
are not kinematically allowed.  One interesting channel is $Z_2
\to Z_1H$, with $H$ a relatively light Higgs boson. Since a
SM-like $HZ_1Z_1$ coupling is expected, the $Z_2\to Z_1H$ decay
can be induced via $Z'-Z$ mixing. In this case the $Z_2Z_1H$
coupling will be proportional to $\sin 2\theta$. Naively, one
could think that this suppression effect may be compensated by the
$(m_{Z_2}/m_{Z_1})^2$ factor arising from the longitudinal
component of the $Z_1$ boson. However, this is not the case
because the factor $m^{-2}_{Z_1}$ gets cancelled with the factor
$m^2_{Z_1}$ coming from the SM-like coupling. Accordingly, the
decay width is given by
\begin{equation}
\Gamma(Z_2\to Z_1H)=m_{Z_2}\frac{\alpha s^2_{2\theta}}{48s^2_W}
\left(\left(1-x^2-y^2\right)^2+12x^2-4x^2y^2\right)\sqrt{\left(1-(x+y)^2\right)\left(1-(x-y)^2\right)},
\end{equation}
with $x=m_{Z_1}/m_{Z_2}$ and $y=m_H/m_{Z_2}$. It can be seen that,
in contrast with the $Z_2\to W^+W^-$ channel, this mode does not
depend on the $m_{Z_2}/m_{Z_1}$ ratio and thus it is insensitive
to the effects of a heavy $Z_2$ boson.\\

\subsection{The decays $ Z_2\to \bar{f}f $}
From the Feynman rules given above, it is straightforward to
obtain the width for the flavor conserving $Z_2\to \bar{f}f$
decay, which is given by
\begin{equation}
\Gamma(Z_2\to \bar{f}f)=m_{Z_2}\frac{\alpha
N^f_C}{3s^2_{2W}}\sqrt{1-\frac{4m^2_f}{m^2_{Z_2}}}
\left(\left|g^f_{VZ'}\right|^2\left(1+\frac{2m^2_f}{m^2_{Z_2}}\right)
+\left|g^f_{AZ'}\right|^2\left(1-\frac{4m^2_f}{m^2_{Z_2}}\right)\right).
\end{equation}
By simplicity we will take the diagonal elements of the mixing
matrix of the quark sector equal to one. As already mentioned, the
leptonic decays of the $Z_2$ boson are expected to be strongly
suppressed. On the contrary, from Table \ref{TABLE} it is evident
that the decay widths to quark pairs have associated an
enhancement factor $1/\sqrt{1-4s^2_W}$, although the decays $Z_2
\to \bar{q}q$, with $q=d$, $s$, and $t$ would be dominated by the
associated vector coupling since the axial one is proportional to
the factor $\sqrt{1-4s^2_W}$ and thus it is negligible.

The decay width for the FCNC  transition $Z_2\to tq'$ can be
written as
\begin{equation}
\Gamma (Z_2\to \bar{t}q'+\bar{q}'t)=\frac{2m_{Z_2}N_C\alpha
\delta^2_L}{3s^2_{2W}}|V^*_{3q'}V_{3t}|^2\left(1-x^2_t\right)\left(1-\frac{1}{2}x^2_t\left(1+x^2_t\right)\right),
\end{equation}
with $q'=u,\,c$ and $x_t=m_t/m_{Z_2}$. The $q'$ mass has been
neglected.

\section{Discussion}
\label{br} One interesting feature of the $331$ model is that the
masses of the new gauge bosons not only can be bounded from below
using the available experimental data, but also from above due to
the constraint $s^2_W<1/4$ imposed by the model. This condition
translates into the upper bound $m_{Z_2}<3.1$ TeV \cite{NG2}.
Though this bound can be relaxed by introducing more complex Higgs
sectors \cite{FP2}, in the following discussion we will assume
that $m_{Z_2}\lesssim 3$ TeV. Even more, using the
symmetry-braking hierarchy $u\gg v_1,v_2,v_3$, this bound can be
translated into an upper bound on the bileptons masses, which is
given by $m_{Y}<m_{Z_2}/2$ \cite{NG2}. This constraint will be
assumed for the bileptons masses, which in addition will be taken
as degenerate. This means that the decays of $Z_2$ into bilepton
pairs would always be kinematically allowed.

Currently the most stringent lower bound on the doubly charged
bilepton arises from muonium-antimuonium conversion, $e^+\mu^-\to
e^-\mu^+$, which yields the limit $m_{Y^{++}}\geq 850$ GeV
\cite{MAM}. It has been argued however that this bound can be
evaded in a more general context since it relies on very
restrictive assumptions \cite{CB}. For instance, the scalar
contributions were not considered in the analysis of Ref.
\cite{MAM} though they may give rise to strong cancellations,
thereby relaxing the bound on the bilepton masses. Another strong
limit, $m_{Y^{++}}> 750$ GeV, arises from fermion pair production
and lepton-flavor violating decays \cite{TULLY}. In addition, the
bound $m_{Y^+}>440$ GeV was derived from limits on the muon decay
width \cite{TULLY2}. We would like to stress that all the above
bounds are model dependent, and so the existence of lighter
bilepton gauge bosons is still allowed. As far as the exotic quark
mass is concerned, the lower bound  $m_{Q}>240$ GeV was derived
from the search of SUSY at the Tevatron. This bound would reach
the level of $320$ GeV at the Run-2 \cite{BEQ}.

We now consider two scenarios which are very illustrative of the
possible behavior of the $Z_2$ boson in the 331 model. Throughout
our analysis we will assume that the bileptons are degenerate and
have a mass $m_{Y}=500$ GeV. We will also assume that the exotic
quarks are degenerate and consider two specific values for their
mass $m_Q$. In the first scenario we will consider that $m_Q=500$
GeV, whereas in the second scenario, we will take $m_Q=m_{Z_2}$.
For the mass of the $Z_2$ boson, the range $1$ TeV $<m_{Z_2}<3$
TeV will be considered. As far as the mixing angle $\theta$ is
concerned, it has been constrained within the context of the 331
model from precision experiments at the $Z$-pole and neutral
current experiments. It was found that $-0.0006<\theta <0.0042$
for $m_{Z_2}>490$ GeV \cite{NG2}. In both scenarios we will use
$\sin\theta \approx \theta =10^{-3}$ and $m_H=120$.

\begin{figure}
 \centering
\includegraphics[width=3.5in]{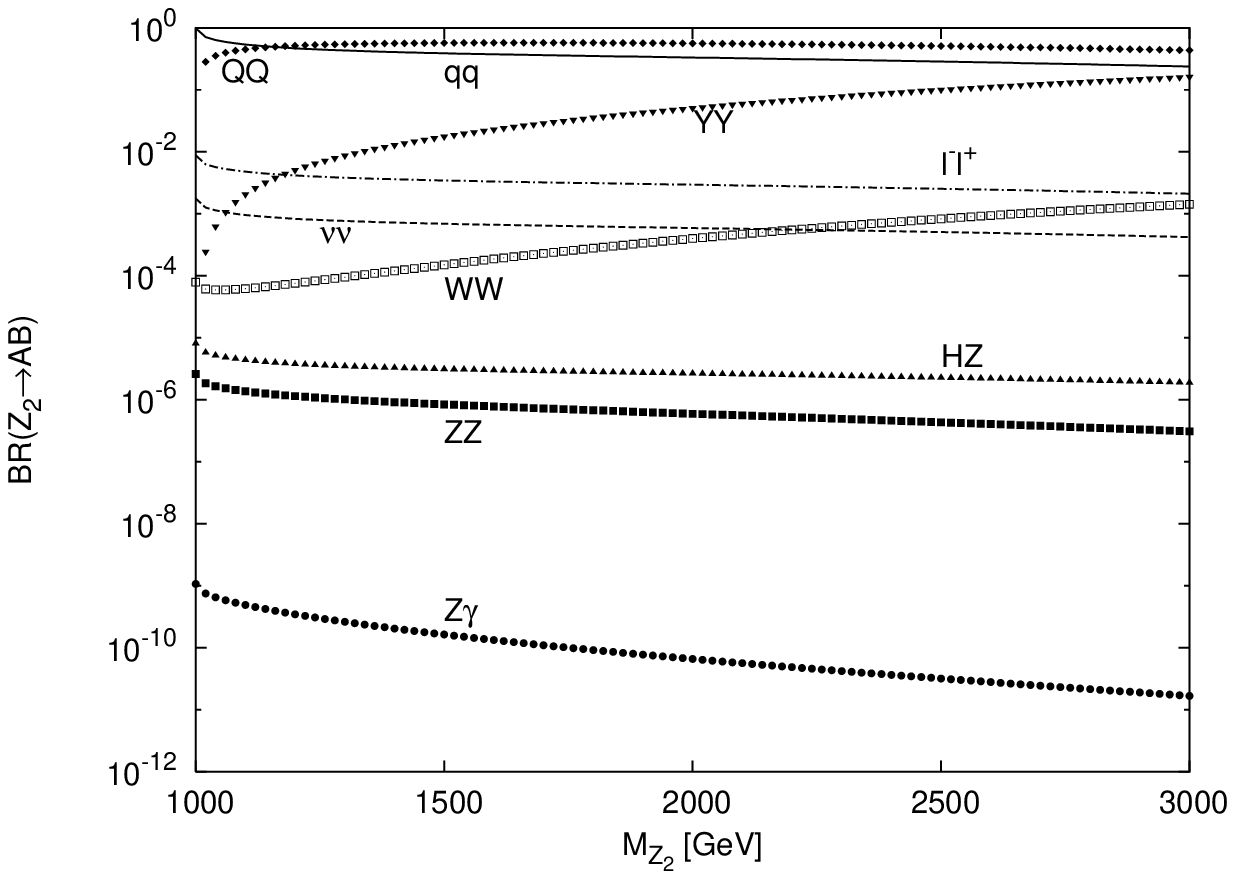}
\caption{\label{GRA1}Branching ratios for the two-body decays of
the $Z_2$ boson as a function of $m_{Z_2}$ in the scenario in
which $m_{Q}=500$ GeV, $m_Y=500$ GeV and $\sin \theta=10^{-3}$.}
 \end{figure}

\begin{figure}
 \centering
 \includegraphics[width=3.5in]{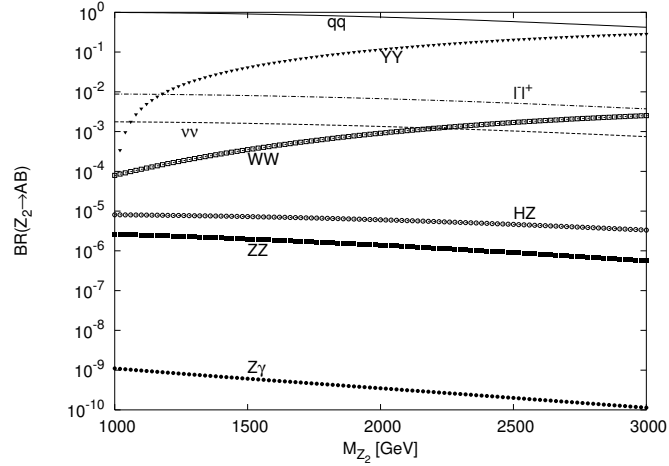}
 \caption{\label{GRA2}Branching ratios for the two-body decays of
the $Z_2$ boson  as a function of $m_{Z_2}$ in the scenario
 in which $m_{Q}=m_{Z_2}$, $m_Y=500$ GeV and $\sin \theta=10^{-3}$.}
 \end{figure}

To begin with, we will discuss the flavor conserving decays of the
$Z_2$ boson, whereas the study of the FCNC transitions will be
deferred as it is necessary to make some assumptions for the
mixing-matrix elements $|V^*_{3q'}V_{3t}|^2$. The branching ratios
for the decay modes $Z_2\to \bar{Q}Q$, $\bar{q}q$, $l^+l^-$,
$\bar{\nu}_l\nu_l$, $YY^\dag$, $WW$, $Z_1H$, $Z_1Z_1$, and
$Z_1\gamma$ are displayed in Fig. \ref{GRA1} as a function of
$m_{Z_2}$ in the scenario in which $m_Q=500$ GeV. In this Fig.,
$Z_2 \to \bar{q} q$ and $Z_2\to \bar{Q}Q$ stand for the decay to
all the exotic quarks and all the SM quarks, respectively. It can
be observed that these modes are dominant, being $Br(Z_2 \to
\bar{Q}Q)$ larger than $Br(Z_2\to \bar{q}q)$. On the contrary, the
$Z_2 \to l^+l^-$ mode is marginal, with a branching ratio of the
order of $10^{-3}$. The invisible mode is still more suppressed,
with $Br(Z_2\to \bar{\nu} \nu)\approx 10^{-4}$. This illustrates
the leptophobic nature of the $Z_2$ boson. As can be noticed,
there is a difference of one order of magnitude between the
branching ratios for the two last modes, which can be explained
from the fact that the vector coupling of the $Z_2$ boson to
charged leptons has associated an additional factor of $3$ (see
Table \ref{TABLE}). As for the $Z_2\to W^+W^-$ mode, its branching
ratio may reach the level of $10^{-3}$ for large values of
$m_{Z_2}$. Since this mode is very sensitive to the mass of the
extra $Z$ boson, it is expected to have a larger branching
fraction if $Z_2$ is heavier. In contrast, $Br(Z_2 \to Z_1H)$ is
of the order of $10^{-5}$ at most and it is almost independent of
the value of the $Z_2$ mass. As far as the anomaly-induced decays
are concerned, $Br(Z_2\to Z_1Z_1)$ is of the order of $10^{-6}$
and is less sensitive to the mass of the extra $Z$ boson than
$Br(Z_2\to Z_1 \gamma)$. This fact can be explained from the fact
that while the $Z_2 \to Z_1 Z_1$ decay width increases with
$m_{Z_2}$, the loop amplitudes decrease in the same proportion
with $m_{Z_2}$. A somewhat different behavior is found for
$Br(Z_2\to Z_1\gamma)$, which is more sensitive to $m_{Z_2}$ and
lies in the range $10^{-9}-10^{-11}$ for the $m_{Z_2}$ values
shown in Fig. \ref{GRA1}. Finally, in the scenario in which
$m_Q=m_{Z_2}$, the respective branching ratios are shown in Fig.
\ref{GRA2}. In this case, the decay to exotic quarks is not
kinematically allowed and thus the main decay channel is $Z_2 \to
\bar{q}q$. In general terms, the branching ratios for all the
remaining decay channels are slightly larger than those shown in
Fig. \ref{GRA1}.

It is interesting to compare our results for the anomaly-induced
decays with those found in Ref. \cite{Dchang}, which were obtained
within the context of a superstring-inspired $E_6$ model. In that
case, it was found that $Br(Z_2\to Z_1Z_1)\sim 10^{-5}$ and
$Br(Z_2\to Z_1\gamma)\sim 10^{-6}$ for a relatively light extra
$Z$ boson with a mass of the order of $0.5$ TeV, though these
branching ratios were obtained by considering the $Z_2 \to e^+e^-$
mode as the dominant one. We believe that our results are
realistic since they were obtained within the context of a model
that generates nontrivial couplings in the neutral current sector,
which in addition depends on known parameters. Though in this
model the $Z_2$ boson has a leptophobic character, it should be
emphasized that these contributions are not important in general
as these decays are only sensitive to heavy fermions. As can be
seen in Table \ref{TABLE}, both the vector and axial couplings of
$Z_2$ to SM and exotic quarks are as important as those existing
in the SM for the $Z_1$ boson. We do not expect that these
branching ratios can be substantially enhanced within the context
of other renormalizable theories.

We now turn to the FCNC decays of the $Z_2$ boson. In order to get
an estimate for the corresponding branching fractions, it is
necessary to assume some value for the unknown coefficient
$|V^*_{3q'}V_{3t}|^2$. In principle, a bound on the $Z_2\,t\,u$
coupling can be indirectly obtained from the $\kappa_{tu\gamma}$
form factor associated with the one-loop vertex $tu\gamma$, for
which the bound $|\kappa_{tu\gamma}|<0.27$ was set recently
\cite{H1}. In addition to the SM contribution, which turns out to
be very suppressed \cite{Soni}, in the 331 model the $t\to
u\gamma$ decay receives contributions from the $Z_2tu$ coupling
and the charged flavor changing currents mediated by the bilepton
gauge bosons and the charged scalars. Both contributions were
analyzed in the context of the $b\to s\gamma$ decay \cite{PHF}.
Unfortunately the resulting bound on $|V^*_{3u}V_{3t}|$ is very
poor due mainly to the large discrepancy between the scales $m_t$
and $m_{Z_2}$. In fact, considering only the $Z_2tu$ contribution
to $\kappa_{tu\gamma}$, the above bound leads to
\begin{equation}
|V^*_{3u}V_{3t}|<15.5\frac{m^2_{Z_2}}{m^2_t},
\end{equation}
which happens to be very poor even for a relatively light $Z_2$
boson. In order to estimate the branching fractions for the FCNC
decays, we will take a different approach, which consists in
assuming that the $V^*_{3u}V_{3t}$ and $V^*_{3c}V_{3t}$
coefficients are of the same order of magnitude as those
associated with the SM one-loop induced couplings $Ztu$ and $Ztc$,
which are proportional to the products of the Kobayashi-Maskawa
matrix elements $V^*_{tb}V_{cb}$ and $V^*_{tb}V_{ub}$. We then
assume the following scenario
\begin{eqnarray}
&&|V^*_{3u}V_{3t}|^2\sim |V^*_{tb}V_{ub}|^2\approx 2.3\times
10^{-5}, \\
&&|V^*_{3c}V_{3t}|^2\sim |V^*_{tb}V_{cb}|^2\approx 1.9\times
10^{-3},
\end{eqnarray}
where the values of the Kobayashi-Maskawa matrix elements were
taken from Ref. \cite{PDG}. To show that this is a reasonable
assumption, let us to consider the FCNC transitions in the down
sector, for which there are already more acceptable bounds
\cite{NG2}. According to our assumption, the respective
coefficients are
\begin{eqnarray}
&&|V^*_{3d}V_{3s}|^2\sim |V^*_{td}V_{ts}|^2\approx 3.8\times
10^{-7}, \\
&&|V^*_{3d}V_{3b}|^2\sim |V^*_{td}V_{tb}|^2\approx 1.9\times
10^{-4}, \\
&&|V^*_{3s}V_{3b}|^2\sim |V^*_{ts}V_{tb}|^2\approx 1.9\times
10^{-3},
\end{eqnarray}
which is to be contrasted with the limits derived from
$K^0-\bar{K}^0$ and $B^0-\bar{B}^0$ mixing \cite{NG2}:
\begin{eqnarray}
&&|V^*_{3d}V_{3s}|<2.5\times 10^{-5}, \\
&&|V^*_{3d}V_{3b}|^2<4\times 10^{-4}.
\end{eqnarray}

Bearing in mind the above assumption, we can estimate the
branching ratios for $Z_2\to tu$ and $Z_2\to tc$. In the limit of
massless $t$, which is a good assumption for the range of values
that we are considering for $m_{Z_2}$, the branching ratio for the
FCNC transitions can be written in terms of that of the $Z_2\to
\bar{t}t$ mode as follows:
\begin{equation}
Br(Z_2\to
\bar{t}q'+\bar{q'}t)=\frac{2|\delta^2_L|V^*_{3q'}V_{3t}|^2}{|g^t_{VZ'}|^2+|g^t_{AZ'}|^2}Br(Z_2\to
\bar{t}t) <3|V^*_{3q'}V_{3t}|^2Br(Z_2\to \bar{t}t).
\end{equation}
In the scenario with $m_{Q}=500$ GeV, $Br(Z_2\to \bar{t}t)\sim
0.1$, so $Br(Z_2\to \bar{t}u+\bar{u}t)\sim 7\times 10^{-6}$ and
$Br(Z_2\to \bar{t}c+\bar{c}t)\sim 6\times 10^{-4}$. The situation
is not very different when $m_{Q}=m_{Z_2}$. In such a scenario
$0.1<Br(Z_2\to \bar{t}t)<0.5$, which yields $7\times
10^{-6}<Br(Z_2\to \bar{t}u+\bar{u}t)< 3.5\times 10^{-5}$ and
$6\times 10^{-4}<Br(Z_2\to \bar{t}c+\bar{c}t)<3\times 10^{-3}$.

\section{Summary}
\label{c} The detection  of an extra $Z$ boson would constitute a
clear evidence that the SM gauge group needs to be extended. The
phenomenology of this particle is rather model dependent and it is
necessary to investigate its physical properties within the
context of some specific models. In this work we have presented a
study on the main decay channels of the extra $Z$ boson predicted
by the minimal $331$ model, including the anomaly-induced decays
$Z_2\to Z_1\gamma$ and $Z_2\to Z_1Z_1$. In the 331 model the
couplings of the extra neutral gauge boson to the fermions and the
extra charged gauge bosons are defined in terms of known
parameters. In particular, the leptophobic and bileptophobic
nature of the $Z_2$ boson is a direct result arising from the
quantum number assignment rather than  an {\it ad hoc} imposition.
These features, together with the fact that the new gauge boson
masses are bounded from above and the unique mechanism of anomaly
cancellation, provide a very interesting scenario to study the
physics of an extra $Z$ boson.

The two-body decay modes $Z_2\to \bar{Q}Q$, $\bar{q}q$, $l^+l^-$,
$\bar{\nu}_l\nu_l$, $YY^\dag$, $W^+W^-$, $Z_1H$, $Z_1Z_1$, and
$Z_1\gamma$ were studied. Also, we outlined some scenarios which
may arise from the existence of tree-level FCNC transitions. In
particular we concentrate on the $Z_1\to tc$ and $Z_2\to tu$ decay
modes, which may be important as the third generation quarks have
a different quantum number assignment. For the purpose of the
numerical analysis, the bileptons were assumed to be degenerate
and their masses were taken in the range $2\,m_{Y}<m_{Z_2}$. As
far as the exotic quark masses, two scenarios were considered: one
in which the decay $Z_2\to \bar{Q}Q$ is kinematically allowed and
another in which these particles have masses of the order of
$m_{Z_2}$. It was found that the $Z_2$ boson would decay
dominantly into SM and exotic quarks, with a combined branching
ratio near to 100 \%. The remaining decay modes can be considered
as rare since are considerably suppressed. In particular, the
leptophobic and bileptophobic nature of the $Z_2$ boson was
discussed. It was found that $Br(Z_2\to YY^\dag)\sim
10^{-2}-10^{-1}$ and $Br(Z_2\to l^+l^-)\sim 10^{-2}$ for $m_{Z_2}$
in the range 1 TeV -- 3 TeV. In the case of the invisible decay
$Z_2\to \bar{\nu}_l\nu_l$, its branching fraction is one order of
magnitude below than that to charged leptons. As far as the
anomaly-induced decays $Z_2\to Z_1Z_1$ and $Z_2\to Z_1\gamma$ are
concerned, they are marginal, with a branching ratio of the order
of $10^{-6}$ and $10^{-9}$, respectively. The tree level FCNC
effects mediated by the $Z_2$ boson were studied in the up sector
of the model. The branching fractions for the $Z_2\to tc$ and
$Z_2\to tu$ decays were estimated by assuming a reasonable
scenario for the elements of the FCNC mixing matrix. It was found
that the corresponding branching ratios can reach the level of
$10^{-3}$ and $10^{-5}$, respectively. Finally, the $Z'-Z$ mixing
induced decay $Z_2\to W^+W^-$ has a branching ratio which is
sensitive to the $Z_2$ mass and may be up to $10^{-2}$ for
$m_{Z_2}=3$ TeV. This decay might be enhanced in models which
allow a very heavy extra $Z$ boson. In contrast, the $Z_2\to Z_1H$
decay has a branching ratio of the order of $10^{-5}$ and is much
less sensitive to a heavier extra $Z$ boson. This decay is
expected to be strongly suppressed.

\acknowledgments{We acknowledge support from Conacyt and SNI (M\'
exico). J.J.T. also acknowledges support from VIEP-BUAP under
grant II 25G02. The work of G.T.V. is partially supported by
SEP-PROMEP.}

\end{document}